\documentclass[superscriptaddress,twocolumn,aps,prl,floatfix,showpacs,amsmath,amssymb]{revtex4}
\usepackage{epsfig}
\usepackage{dcolumn}
\usepackage{bm}
\begin{document}
\title{
Evidence for Charge Glass-like Behavior in Lightly Doped La$_{2-x}$Sr$_{x}$CuO$_{4}$ at Low Temperatures}

\author{I. Rai\v{c}evi\'{c}}\email{raicevic@magnet.fsu.edu}
  \affiliation{National High Magnetic Field Laboratory, Florida State University, Tallahassee,
FL 32310, USA}
\affiliation{Department of Physics, Florida State University, Tallahassee,
FL 32306, USA}
\author{J. Jaroszy\'{n}ski}
 \affiliation{National High Magnetic Field Laboratory, Florida State University, Tallahassee,
FL 32310, USA}
\author{Dragana Popovi\'c}
\affiliation{National High Magnetic Field Laboratory, Florida State University, Tallahassee,
FL 32310, USA}
\affiliation{Department of Physics, Florida State University, Tallahassee,
FL 32306, USA}
\author{C. Panagopoulos}
\affiliation{Cavendish Laboratory, University of Cambridge,
Cambridge CB3 OHE, United Kingdom}
\affiliation{Department of Physics, University of Crete and FORTH, 71003 Heraklion, Greece}
\author{T. Sasagawa}
\affiliation{Materials and Structures Laboratory, Tokyo
Institute of Technology, Kanagawa 226-8503, Japan }
\date{\today}

\begin{abstract}
A $c$-axis magnetotransport and resistance
noise study in La$_{1.97}$Sr$_{0.03}$CuO$_{4}$ reveals clear signatures of glassiness, such as hysteresis, memory, and slow, correlated dynamics, but only at temperatures ($T$) well below the spin glass transition temperature $T_{sg}$.
The results strongly suggest the emergence of charge glassiness, or dynamic charge ordering, as a result of Coulomb interactions.

\end{abstract}
\pacs{74.72.Dn, 72.70.+m, 75.50.Lk}


\maketitle

The role of heterogeneities observed in most hole-doped
high-temperature superconductors (HTS) is one of the major open
issues in the field \cite{Millis,Elbio}.  In weakly doped Mott
insulators, such as HTS, charge heterogeneities are expected to arise
due to the existence of several competing ground
states \cite{nanoscale-phasesep}, and may be even to exhibit glassy dynamics \cite{Schmalian}.
Even a small amount of disorder may favor glassiness over various static charge-ordered states \cite{Pankov-scr}.
Experiments in hole-doped HTS suggest \cite{NQR,IR,padilla05,Christos-Vlad} that
glassiness of both spins and charges emerges with the first added holes and evolves with doping $x$.
While spin glass (SG) behavior is well established at low $T$, the evidence for glassy freezing of charges
is not conclusive.  Hence, alternative, bulk probes of charge dynamics are needed to explore the nature of the ground state.  We present a novel study of the charge dynamics in a lightly doped La$_{2-x}$Sr$_{x}$CuO$_{4}$ (LSCO) using a combination of transport and noise spectroscopy that proved to be a powerful probe of dynamics in other glasses
\cite{weissman93,noise-Si}.  We find several clear signatures of glassiness at $T\ll T_{sg}$.
The data strongly suggest that the doped holes form a \textit{dynamically} ordered, cluster glass
as a result of Coulomb interactions.

\begin{figure}
\includegraphics[width=7.3cm]{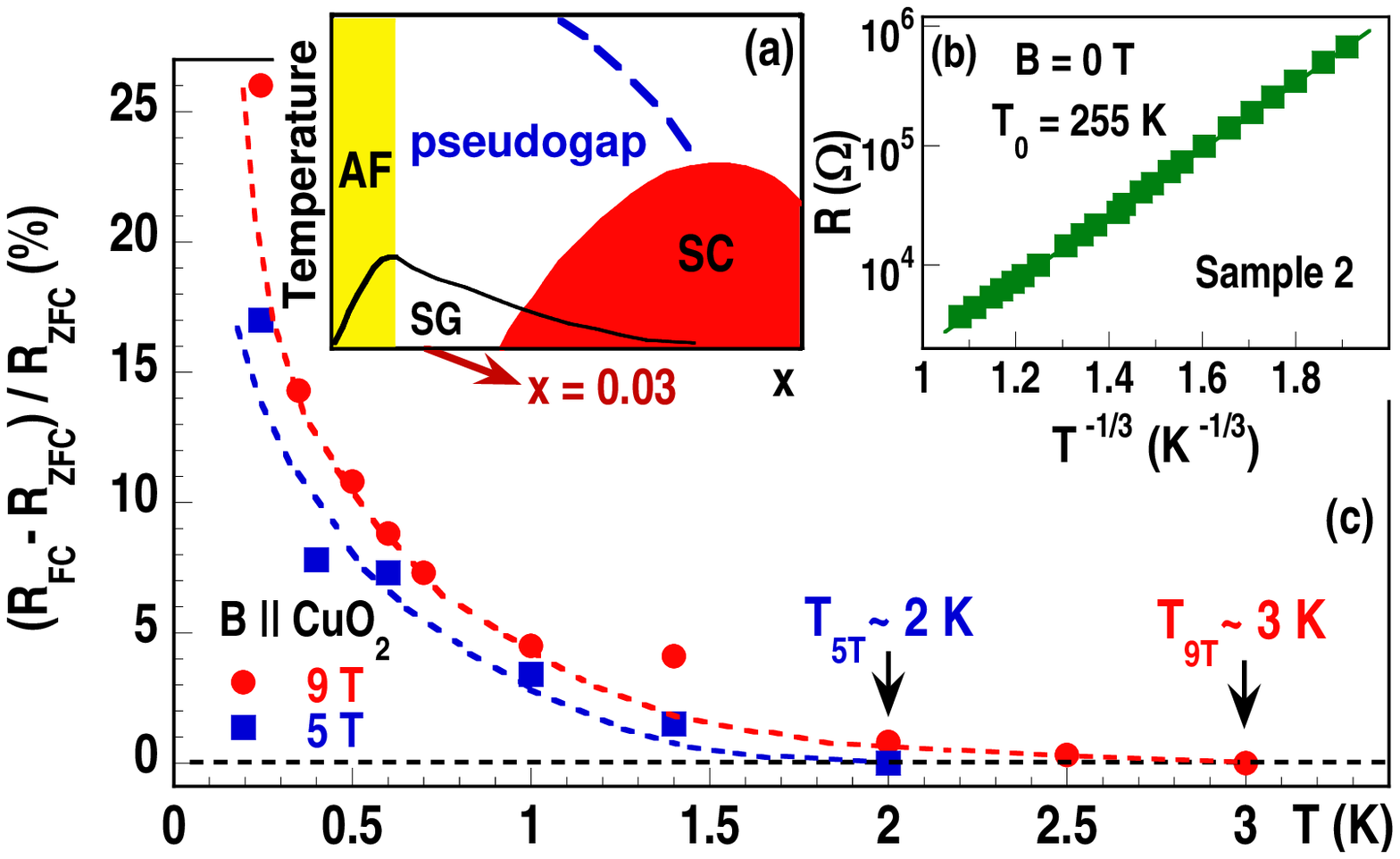}
\caption {(color online)  (a) Schematic phase diagram of hole-doped cuprates.
(b) ZFC $R(T)$.  (c) Sample 2.  The differences between field-cooled (FC)
(squares: in 5 T, dots: in 9 T) and ZFC $R$ \textit{vs.} $T$ vanish at $T_{B}=T_{5\textrm{\scriptsize T}}$ and $T_{9\textrm{\scriptsize T}}$, respectively.  Dashed lines guide the eye.  The same results were obtained with 0.24,
0.014, and 0.007 K/min cooling rates.
}\label{fig:ZFCvsFC}
\end{figure}

In LSCO, the prototypical cuprate HTS, the three-dimensional (3D) long range antiferromagnetic (AF) order of the parent compound is destroyed above $x\approx 0.02$, but 2D short range AF correlations persist~\cite{Kastner-98review}.
In particular, as a result of hole doping, CuO$_2$ (\textit{ab}) planes develop a pattern of AF domains that are separated by antiphase boundaries~\cite{Matsuda,magsusc-Wakimoto,magsusc-Lavrov}.  Since the Dzyaloshinskii-Moriya interaction induces slight canting of the spins in CuO$_2$ planes towards the $c$ axis,
there is a weak ferromagnetic (FM) moment in the \textit{bc} plane associated with each AF domain, such that the direction of the FM moment is uniquely linked to the phase of the AF order~\cite{Thio,magsusc-Lavrov}.
The interplane exchange favors staggered ordering of those FM moments in the $c$ direction.  At low enough
$T<T_{sg}(x)$, the system freezes into a SG \cite{Kastner-98review,magsusc-Wakimoto,magsusc-Lavrov} that
extends into the superconducting (SC) phase for $x>0.05$ \cite{Niedermayer} up to optimal doping \cite{Christos2002} [Fig.~\ref{fig:ZFCvsFC}(a)].
Various experiments  on lightly doped LSCO (\textit{e.g.} Refs. \cite{NQR,IR}), including transport studies \cite{Ando},
were interpreted in terms of the hole-poor AF domains separated by the hole-rich regions in CuO$_2$ planes,
with infrared studies being inconsistent with the notion of static charge ordering \cite{padilla05}.

We report an extensive
study of the low-$T$ ($T\sim 1$~K and below) $c$-axis magnetotransport and low-frequency resistance ($R$) noise in LSCO with $x=0.03$ [Fig.~\ref{fig:ZFCvsFC}(a)].  Such lightly doped samples are insulating at low $T$ and, hence, most likely to exhibit charge glassiness \cite{eglass}.
The high-quality single crystal was grown by the traveling-solvent floating-zone
technique \cite{sasagawa98}; $T_{sg}\sim 7$\,-$8$~K \cite{Christos-Vlad}.  Two samples were cut along
the main crystallographic axes and polished into
$0.6 \times 0.8 \times 1.57$ mm$^{3}$ (sample 1) and $0.6 \times 0.9 \times
1.6$ mm$^{3}$ (sample 2) bars
suitable for direct $c$-axis transport measurements.  Both samples had the same behavior.  $R$ was measured
with a standard four-probe ac method ($\sim 7$~Hz) in the Ohmic regime, and noise with a five-probe ac bridge method
\cite{Sco87}, using a lock-in amplifier
($\sim 7$~Hz; excitation current $I_{exc}=1$~nA in the
Ohmic regime) to detect the difference voltage.  This method
minimizes the influence of $T$ and $I_{exc}$ fluctuations on the
$R$ noise.  Additional care was taken to
ensure that the observed noise did not come from the contacts.  In
particular, the noise was measured twice on sample 2. For the
first run, the contacts were made using the Dupont $6838$ Ag paste
and, for the second run, by evaporating Au.
In both cases, the room $T$ contact resistances were
$<1~\Omega$, and the noise characteristics were the same.

The $c$-axis $R(T)$ exhibits insulating behavior, which obeys the
variable-range hopping $R=R_0\exp(T_0/T)^{\mu}$ for $T\leq
1$~K [Fig.~\ref{fig:ZFCvsFC}(b)].  The best fit to the data is
obtained with $\mu = 1/3$, consistent with early results on
ceramic LSCO samples, where $\mu=1/2$ for $x=0.02$ and $\mu=1/4$
for $x=0.05$~\cite{LSCO-hopping}.  Similar doping dependence of
$\mu$ has been observed in various systems, such as doped
semiconductors~\cite{ES-book} and other disordered Mott
insulators~\cite{Satoru}.  Surprisingly, here the
precise form of $R(T, B=0)$ depends on the cooling protocol. In
particular, at low $T$, $R(B=0)$ obtained after cooling in field
$B_{FC}$ is higher than the zero-field cooled (ZFC) $R(B=0)$.
This difference decreases with increasing $T$, and vanishes at a
temperature $T_B$ that grows with $B_{FC}$  [Fig.~\ref{fig:ZFCvsFC}(c)].
This history dependence, with nearly the same
magnitude and $T_B$, was observed for both $B\parallel c$ and
$B\perp c$.

Strong history dependent effects are seen
also if $B$ is applied after zero-field cooling.
\begin{figure}
\includegraphics[clip,width=7.5cm]{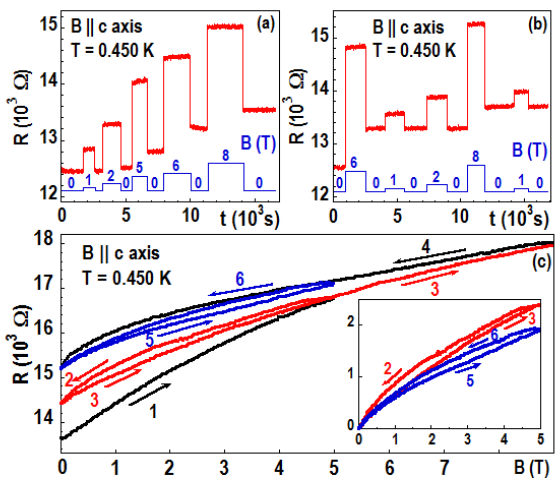}
\caption {(color online)  Sample 1.  $R$ \textit{vs.} time $t$ upon the subsequent
application and removal of (a) $B = 1, 2, 5, 6, 8$~T, and (b) $B = 6, 1,
2, 8, 1$~T.  (c) Sample 2; $R$ \textit{vs.} $B$. The arrows and numbers denote the direction and the order of $B$-sweeps.  The sweep rate (0.001~T/min for $B<1$~T, 0.005~T/min for $B>1$~T) was low enough to avoid the sample heating.  Inset: Subloops shifted vertically to 0 for comparison.  The hysteretic and memory effects are similar for
positive and negative $B$, and for both
$B\parallel c$ and $B\perp c$.  They become smaller with increasing $T$, and vanish for $T\gtrsim 1$~K.}\label{fig:MEMORYandHYSTERESIS}
\end{figure}
%
Figure~\ref{fig:MEMORYandHYSTERESIS}(a) shows $R$ as a function of time $t$
upon the subsequent application and removal of several $B$ values at a fixed, low $T$.
Obviously, $R$ increases with $B$.  However, whenever $B$ is
turned off, $R$ decreases, but it remains higher than the  previous $R(B=0)$.
In fact, $R(B=0)$ keep increasing as long as subsequent $B$ increase monotonically.
Otherwise [Fig.~\ref{fig:MEMORYandHYSTERESIS}(b)], $R(B=0)$ is determined by
the highest previous $B$: the sample acquires a \textit{memory} of its magnetic history.
Similar memory effects in transport have been seen in other systems, such as manganites \cite{levy02} in the regime of phase separation and heavily underdoped YBa$_2$Cu$_3$O$_{6+x}$ (YBCO) at low $T$ \cite{andoYBCO99}.
In YBCO, they were attributed to the freezing of the directionally ordered charge stripes.
In LSCO, the hysteretic and memory effects may result from the irreversible orienting of the FM moments \cite{shapememory-comment} in the direction of the applied $B$.  Therefore, it may be instructive
to study the hysteresis in more detail.

In order to obtain the data with a controlled history [\textit{e.g.} Figs.~\ref{fig:MEMORYandHYSTERESIS}(a)-(c)] the samples' memory was first erased by warming up above $T\sim 1$~K ($<T_{sg}$), where hysteretic effects vanish \cite{warming-comment}.
The subsequent cooling to the measurement $T$ produced small $R(t)$ relaxations only at the lowest $T\sim 0.1$~K, but any possible intrinsic nonequilibrium dynamics could not be separated out from the effects of cooling.
The ZFC magnetoresistance (MR) $R(B)$ [Fig.~\ref{fig:MEMORYandHYSTERESIS}(c)] first follows the paths 1-3 during the cycling of $B$ between 0 and 5~T.  After the second sweep to 5~T, the first
subloop (2-3) has closed, and continuing to raise $B$ does not disturb the structure of the outer loop.
The system thus exhibits return-point memory (RPM).  The second closed subloop (5-6), obtained between the
same $B$ end points but with a different history, is clearly incongruent with the first one [Fig.~\ref{fig:MEMORYandHYSTERESIS}(c) inset].  This hysteresis in MR is strikingly reminiscent of the behavior of magnetization in various magnetic materials~\cite{Bertotti}, including spin glasses~\cite{Binder-Young}.  This suggests the existence of magnetic domains (clusters, or switching units) with holes confined within the domain walls,
supported by the results below and consistent with earlier studies \cite{Matsuda,magsusc-Wakimoto,magsusc-Lavrov,NQR,IR,Ando,padilla05}.
The novel observations of the RPM and the incongruence of the subloops, which indicates that domains interact, impose strong constraints on theory.
A model of an electron nematic in CuO$_{2}$ planes \cite{kivelson06}
finds the same type of hysteresis, but in the in-plane $R$ anisotropy, so it is unclear if that can be related to our work.

\begin{figure}
\includegraphics[width=7.5cm]{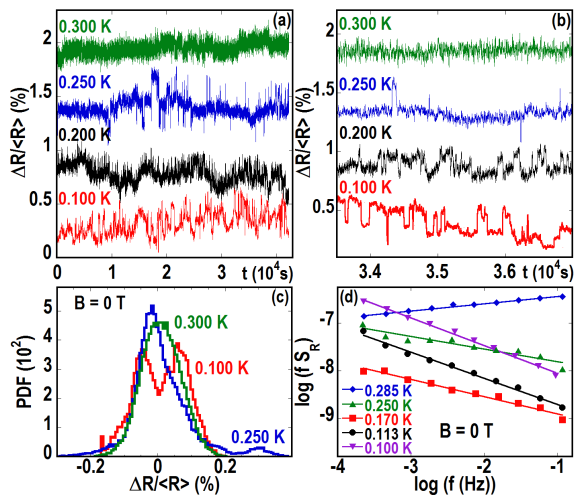}
\caption {(color online) Sample 1.  (a) $\Delta R/\langle R
\rangle =(R-\langle R \rangle)/\langle R \rangle$ ($\langle R
\rangle$ -- the time-averaged $R$) \textit{vs.} $t$ for $B=0$ at several $T$.  (b) An
expanded section of (a).  All traces are shifted for clarity.  (c) PDF
$vs$. $\Delta R/\langle R\rangle$ for the 12-hour time interval.
(d) The octave-averaged power spectra $S_{R}(f)$
have been corrected for the white background noise.  Solid lines are fits to $S_{R}\propto 1/f^{\alpha}$.
}\label{fig:NOISEinZEROB}
\end{figure}
\begin{figure}
\includegraphics[width=7.5cm]{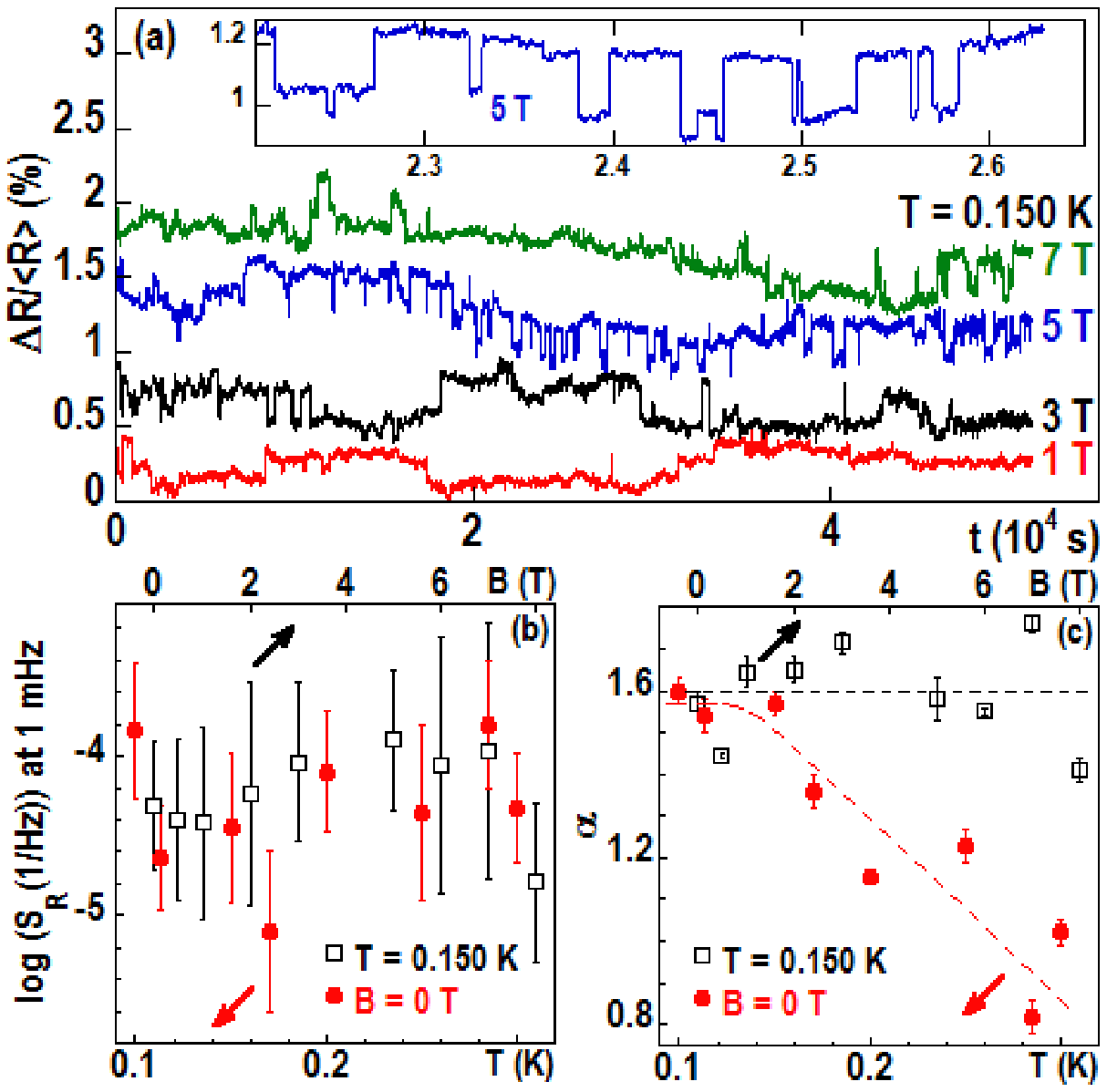}
\caption {(color online) Sample 1.  (a) ZFC $\Delta R/\langle R
\rangle$ \textit{vs.} $t$ at $T=0.150$~K for several $B\parallel
c$. All traces are shifted for clarity.  Inset: A
section of the noise at 5~T.  (b) $S_{R} (f=1$~mHz) and (c)
$\alpha$ \textit{vs.} $T$ ({\large $\bullet$}) and $B$ ({\scriptsize $\square$}).
The error bars in (b) are standard deviations of the
data.  Dashed lines in (c) guide the eye.}\label{fig:NOISEinDIFFB}
\end{figure}

The noise, \textit{i.e.} the time series of the relative changes in resistance $\Delta R(t)/\langle R
\rangle$, also provides evidence for the existence of some switching units.
Figures~\ref{fig:NOISEinZEROB}(a) and \ref{fig:NOISEinZEROB}(b)  show clearly that, at low $T$, the system exhibits switching between well distinguishable states with the characteristic lifetimes
ranging from a few minutes to several hours.  Switching noise was also observed in YBCO, on much shorter
$t$ scales and at much higher $T$, in the pseudogap regime~\cite{Bonetti}.  It is striking that here the noise amplitude is comparable to that in YBCO, even though our sample volume is about $10^{11}$ times larger.
Also, here the switching noise is superposed on other random, slow fluctuations.
The histograms of $\Delta R/\langle R \rangle$
values [Fig.~\ref{fig:NOISEinZEROB}(c)] confirm that, at the lowest $T$ (\textit{e.g.} $T\sim 0.1$~K), the system appears to prefer two states, but other states are also present.  Moreover, the precise shape of the probability density function (PDF) depends randomly on the observation time (not shown), signaling that the system is nonergodic (glassy) on experimental time scales.  All these effects become less prominent as $T$ is raised, and the PDF becomes Gaussian already by
0.3~K.  The noise power spectra $S_R(f)$ ($f$ -- frequency) obey $S_{R}\propto 1/f^{\alpha}$
[Fig.~\ref{fig:NOISEinZEROB}(d)].  In order to compare the noise magnitudes at different $T$ and $B$,
$S_{R}(f=1$~mHz) is taken as the measure of noise, and it is determined from the fits in Fig.~\ref{fig:NOISEinZEROB}(d).  $S_{R}(f=1$~mHz) does not depend on $T$ [Fig.~\ref{fig:NOISEinDIFFB}(b)], as is already apparent from the variance of the raw data [Fig.~\ref{fig:NOISEinZEROB}(a)].  However, $\alpha$ increases from $\sim 1$ to $\sim 1.6$ as $T$ is reduced from 0.3~K to 0.1~K [Fig.~\ref{fig:NOISEinDIFFB}(c)], reflecting the slowing down of the dynamics and the increasing non-Gaussianity of the noise, similar to other systems out of equilibrium~\cite{noise-Si}.

The noise was measured also in $B\parallel c$ of up to 9~T.
Surprisingly, $B$ does not seem to have any effect on the noise.
At low $T$, both switching events and other fluctuations on many
different time scales are still present
[Fig.~\ref{fig:NOISEinDIFFB}(a)], the noise magnitude is
independent of $B$ [Fig.~\ref{fig:NOISEinDIFFB}(b)], and
$S_{R}\propto1/f^{\alpha}$ with $\alpha\sim 1.6$ that remains
unchanged by $B$ [Fig.~\ref{fig:NOISEinDIFFB}(c)].
Furthermore, we have established that, unlike $R$ itself
(Figs.~\ref{fig:ZFCvsFC} and ~\ref{fig:MEMORYandHYSTERESIS}), all the
noise characteristics are independent of the
magnetic history.

\begin{figure}
\includegraphics[width=6.8cm]{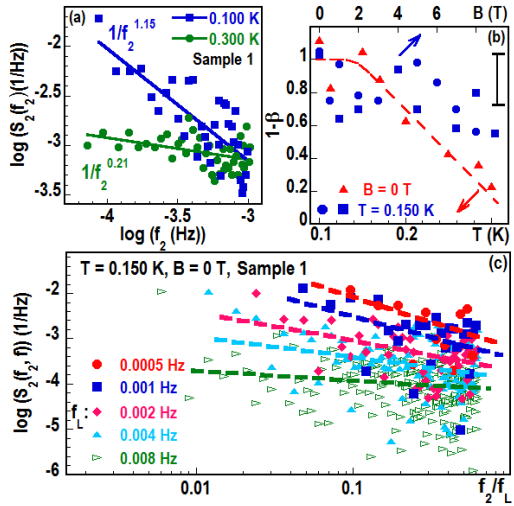}
\caption {(color online) (a) The normalized second spectra $S_{2}$($f_{2}$), with the Gaussian background subtracted,
for two $T$.  Solid lines are fits to
$S_2\propto 1/f_{2} ^ {1-\beta}$. (b) (1 - $\beta$) $vs$. $T$ ($\blacktriangle$; sample 1) and $B$ ({\large $\bullet$}: sample 1, {\scriptsize $\blacksquare$}: sample 2).  The dashed line guides the eye.
The results shown
in (a) and (b) were obtained by averaging over the 0.5--1, 1--2, and 2--4 mHz octaves.  The error bar in (b) shows the maximum standard deviation of the data.  (c) $S_{2}$ measured in
octaves $f=(f_{L},2f_{L}$).  Dashed lines are linear fits to guide the eye.\vspace{-4pt}}\label{fig:SSandSCALING}
\end{figure}

The second spectrum $S_{2}(f_{2},f)$, which is the power spectrum of the fluctuations of $S_{R}(f)$ with $t$, probes the correlations between fluctuators: it is white (independent of $f_2$) for uncorrelated fluctuators (Gaussian noise), and $S_{2}(f_{2},f)\propto 1/f_{2}^{1-\beta}$ for interacting ones \cite{weissman93,duttahorn81,seidler93}.
$S_2$ was calculated for a few
octaves $f = (f_{L},2f_{L}$) [Fig.~\ref{fig:SSandSCALING}(a)].  The results [Fig.~\ref{fig:SSandSCALING}(b)] show
clearly an increase of ($1-\beta$) from $\approx 0$ at 0.3~K to large nonwhite values as $T$ is reduced, demonstrating that fluctuators become strongly correlated with decreasing $T$.  On the other hand, $B$ has no effect on large low-$T$ values of ($1-\beta$), \textit{i.e.} $B$ does not seem to affect the nature of the correlations.

As in
studies of other glasses~\cite{weissman93,noise-Si}, we explore the scaling of $S_{2}(f_{2},f)$ with respect to $f_{2}$ and $f$ in order to distinguish hierarchical pictures from
generalized models of interacting, compact droplets or clusters.
In the latter, the
low-$f$ noise comes from a smaller number of large elements because they are slower, while the higher-$f$ noise comes from a larger number of smaller elements that
are faster \cite{weissman93}.  In the presence of short-range interactions, big elements are more likely to interact than small ones and, hence, non-Gaussian effects and $S_2$ will be stronger for lower $f$. 
Indeed, we find that, unlike some well-known spin (\textit{e.g.} CuMn)~\cite{weissman93} or Coulomb~\cite{noise-Si} glasses, $S_{2}(f_{2},f)$ is \textit{not} scale invariant, but rather it decreases with $f$ at constant $f_2/f$ [Fig.~\ref{fig:SSandSCALING}(c)], consistent with droplet models \cite{droplets}.  This result strongly supports
the picture of spatial segregation of holes into locally ordered, \textit{interacting} regions, as inferred from the MR hysteresis, resulting from competing interactions on different
length scales.

The noise study indicates that, at low $T$, the system wanders collectively between many metastable states.  The apparent insensitivity of the noise statistics to $B$ and to the magnetic history strongly suggests that the observed glassiness reflects the slow dynamics of charge, not spins.  In conventional spin glasses,
by contrast, all $R$ noise characteristics are affected by $B$ \cite{Jan-SG}.  In addition,
unlike spin glasses,
here the onset of glassiness in both transport and noise occurs at $T\ll T_{sg}$, providing further evidence that the two phenomena are not directly related.  In fact,
a gradual enhancement of the correlated behavior with decreasing $T$ suggests, in analogy with 2D Coulomb glasses \cite{noise-Si}, that the charge glass transition $T_{cg}=0$.  The noise reveals glassy dynamics only at $T<0.3$~K, consistent with the higher $T$ observations of static, as opposed to dynamic, charge ordering in underdoped cuprates in other experiments on comparable time scales \cite{kohsaka07}.

We have demonstrated that the $c$-axis transport and $R$ noise in $x=0.03$ LSCO exhibit several clear signatures of glassiness, such as hysteresis, memory, and slow, correlated dynamics, as $T\rightarrow 0$.  The data strongly suggest that the doped holes form a cluster glass state as a result of Coulomb interactions \cite{Schmalian}, albeit in the presence of a random, internal magnetic field that is produced by spin freezing.  Further work at higher $x$ is needed to determine how this dynamic charge order evolves into and coexists with the SC state.  Studies of other cuprates, including those grown epitaxially, would be also of interest to see if the emergence of a charge glass is universal to cuprates, and perhaps even to all doped Mott insulators.

We thank G. Jelbert for technical assistance, V. Dobrosavljevi\'c and M. B. Weissman for stimulating discussions,
NSF No. DMR-0403491, NHMFL via NSF No. DMR-0084173, The Royal Society, the EURYI scheme, and MEXT-CT-2006-039047 for financial support.


\begin{thebibliography}{99}
\expandafter\ifx\csname
natexlab\endcsname\relax\def\natexlab#1{#1}\fi
\expandafter\ifx\csname bibnamefont\endcsname\relax
  \def\bibnamefont#1{#1}\fi
\expandafter\ifx\csname bibfnamefont\endcsname\relax
  \def\bibfnamefont#1{#1}\fi
\expandafter\ifx\csname citenamefont\endcsname\relax
  \def\citenamefont#1{#1}\fi
\expandafter\ifx\csname url\endcsname\relax
  \def\url#1{\texttt{#1}}\fi
\expandafter\ifx\csname
urlprefix\endcsname\relax\def\urlprefix{URL }\fi
\providecommand{\bibinfo}[2]{#2}
\providecommand{\eprint}[2][]{\url{#2}}

\bibitem{Millis} J. Orenstein and A. J. Millis, Science \textbf{288}, 468 (2000).

\bibitem{Elbio} E. Dagotto, Science \textbf{309}, 257 (2005).

\bibitem{nanoscale-phasesep} L. P. Gor'kov \textit{et al.},
JETP Lett. \textbf{46}, 420 (1987); S. A. Kivelson, \textit{et al.},
Rev. Mod. Phys. \textbf{75}, 1201 (2003); E. Dagotto, \emph{Nanoscale phase separation and colossal magnetoresistance} (Springer-Verlag, Berlin, 2002).

\bibitem{Schmalian} J. Schmalian \textit{et al.},
Phys. Rev. Lett. \textbf{85}, 836 (2000).

\bibitem{Pankov-scr} S. Pankov \textit{et al.},
Phys. Rev. Lett. \textbf{94}, 046402 (2005).

\bibitem{NQR} M.-H. Julien \textit{et al.}, Phys. Rev. Lett. \textbf{83}, 604 (1999); P. M. Singer \textit{et al.}, Phys. Rev. Lett. \textbf{88}, 047602 (2002).

\bibitem{IR} M. Dumm \textit{et al.},
Phys. Rev. Lett. \textbf{91}, 077004 (2003).

\bibitem[{\citenamefont{Padilla et~al.}(2005)}]{padilla05}
\bibinfo{author}{\bibfnamefont{W.~J.} \bibnamefont{Padilla}}
  \bibnamefont{\textit{et~al}.}, \bibinfo{journal}{Phys. Rev. B}
  \textbf{\bibinfo{volume}{72}}, \bibinfo{eid}{205101} (\bibinfo{year}{2005}).

\bibitem{Christos-Vlad} C. Panagopoulos \textit{et al.},
Phys. Rev. B \textbf{72}, 014536 (2005), and references therein.

\bibitem{weissman93} M. B. Weissman \textit{et al.}, J. Magn. Magn. Mater. \textbf{114}, 87 (1992); M. B. Weissman, Rev. Mod. Phys. \textbf{65}, 829 (1993).

\bibitem[{noi()}]{noise-Si}
\bibinfo{note}{S. Bogdanovich \textit{et al.},
Phys. Rev. Lett. {\bf 88}, 236401 (2002); J. Jaroszy\'nski
\textit{et al.},
Phys. Rev. Lett. {\bf 89}, 276401 (2002);
\textbf{92}, 226403
(2004)}.

\bibitem{Kastner-98review} M. A. Kastner \textit{et al.}, Rev. Mod. Phys. \textbf{70}, 897 (1998).

\bibitem{Matsuda} M. Matsuda \textit{et al.},
Phys. Rev. B \textbf{62}, 9148 (2000);
\textbf{65}, 134515 (2002).

\bibitem{magsusc-Wakimoto} S. Wakimoto \textit{et al.},
Phys. Rev. B \textbf{62}, 3547 (2000).

\bibitem{magsusc-Lavrov} A. N. Lavrov \textit{et al.},
Phys. Rev. Lett. \textbf{87}, 017007 (2001).

\bibitem{Thio} T. Thio \textit{et al.}, Phys. Rev. B \textbf{38}, 905 (1988); \textbf{41}, 231 (1990).

\bibitem{Niedermayer} C. Niedermayer \textit{et al.},
Phys. Rev. Lett. \textbf{80}, 3843 (1998).

\bibitem{Christos2002} C. Panagopoulos \textit{et al.},
Phys. Rev. B \textbf{66}, 064501 (2002).

\bibitem{Ando} Y. Ando \textit{et al.},
Phys. Rev. Lett. \textbf{88}, 137005 (2002);
\textbf{90}, 247003 (2003).

\bibitem[{egl()}]{eglass}
\bibinfo{note}{J.~H. Davies \textit{et al.}, Phys. Rev.
Lett. {\bf 49}, 758 (1982); M. Gr\"{u}newald \textit{et al.}, J. Phys. C~{\bf 15}, L1153 (1982); M. Pollak
\textit{et al.}, Sol. Energy Mater.~{\bf 8}, 81 (1982)}.

\bibitem{sasagawa98} T. Sasagawa \textit{et al.},
Phys. Rev. Lett. \textbf{80}, 4297 (1998).

\bibitem{Sco87} J. H. Scofield, Rev. Sci. Instrum. \textbf{58}, 985 (1987).

\bibitem{LSCO-hopping} B. Ellman \textit{et al.}, Phys. Rev. B \textbf{39}, 9012 (1989).

\bibitem{ES-book} B. I. Shklovskii and A. L. Efros, \textit{Electronic Properties of Doped Semiconductors} (Springer-Verlag, Berlin, 1984).

\bibitem{Satoru} S. Nakatsuji \textit{et al.}, Phys. Rev. Lett. \textbf{93}, 146401 (2004).

\bibitem[{\citenamefont{Levy et~al.}(2002)}]{levy02}
\bibinfo{author}{\bibfnamefont{P.}~\bibnamefont{Levy}} \bibnamefont{\textit{et~al}.},
  \bibinfo{journal}{Phys. Rev. Lett.} \textbf{\bibinfo{volume}{89}},
  \bibinfo{pages}{137001} (\bibinfo{year}{2002}).

\bibitem[{\citenamefont{Ando et~al.}(1999)}]{andoYBCO99}
\bibinfo{author}{\bibfnamefont{Y.}~\bibnamefont{Ando}} \bibnamefont{\textit{et~al}.},
  \bibinfo{journal}{Phys. Rev. Lett.} \textbf{\bibinfo{volume}{83}},
  \bibinfo{pages}{2813} (\bibinfo{year}{1999}).

\bibitem{shapememory-comment} They cannot be due to the ``magnetic shape memory'', since the applied $B$ are too small to cause a swapping of the orthorhombic $a$ and $b$ axes~\cite{Ando-shapememory}.

\bibitem{Ando-shapememory} A. N. Lavrov \textit{et al.},
Nature \textbf{418}, 385 (2002); S. Ono \textit{et al.},
Phys. Rev. B \textbf{70}, 184527 (2004).

\bibitem{warming-comment}  In practice, most of the data were taken after warming up to 10~K ($>T_{sg}$),
obtaining the same results.

\bibitem{Bertotti} G. Bertotti, \textit{Hysteresis in Magnetism} (Academic Press, New York, 1998).

\bibitem{Binder-Young} K. Binder \textit{et al.},
Rev. Mod. Phys. \textbf{58}, 801 (1986).

\bibitem[{\citenamefont{Carlson et~al.}(2006)}]{kivelson06}
\bibinfo{author}{\bibfnamefont{E.~W.} \bibnamefont{Carlson}}
  \bibnamefont{\textit{et~al}.}, \bibinfo{journal}{Phys. Rev. Lett.}
  \textbf{\bibinfo{volume}{96}}, \bibinfo{eid}{097003} (\bibinfo{year}{2006}).

\bibitem{Bonetti} J. A. Bonetti \textit{et al.},
Phys. Rev. Lett. \textbf{93}, 087002 (2004).

\bibitem[{\citenamefont{Dutta and Horn}(1981)}]{duttahorn81}
\bibinfo{author}{\bibfnamefont{P.}~\bibnamefont{Dutta}} \bibnamefont{and}
  \bibinfo{author}{\bibfnamefont{P.~M.} \bibnamefont{Horn}},
  \bibinfo{journal}{Rev. Mod. Phys.} \textbf{\bibinfo{volume}{53}},
  \bibinfo{pages}{497} (\bibinfo{year}{1981}); \bibinfo{note}{{M}. B. Weissman,
  Rev. Mod. Phys. {\bf 60}, 537 (1988)}.

\bibitem{seidler93} G. T. Seidler \textit{et al.}, Phys. Rev. B \textbf{53}, 9753 (1996);
K. M. Abkemeier, Phys. Rev. B \textbf{55}, 7005 (1997).

\bibitem{droplets} D. S. Fisher \textit{et al.}, Phys. Rev. B \textbf{38}, 373 (1988); \textbf{38}, 386 (1988).

\bibitem{Jan-SG} N. E. Israeloff \textit{et al.}, Phys. Rev. Lett. \textbf{63}, 794 (1989); J. Jaroszy\'nski \textit{et al.}, Phys. Rev. Lett. \textbf{80}, 5635 (1998); G. Neuttiens \textit{et al.}, Phys. Rev. B \textbf{62}, 3905 (2000).

\bibitem[{\citenamefont{Kohsaka et~al.}(2007)}]{kohsaka07}
\bibinfo{author}{\bibfnamefont{Y.}~\bibnamefont{Kohsaka}} \bibnamefont{\textit{et~al}.},
  \bibinfo{journal}{Science} \textbf{\bibinfo{volume}{315}},
  \bibinfo{pages}{1380} (\bibinfo{year}{2007}).

\end{thebibliography}
\end{document}